\documentclass[aps,preprint,nofootinbib,superscriptaddress]{revtex4-2}
\usepackage{amsmath}
\usepackage{graphicx}
\usepackage{color}
\usepackage{slashed}
\usepackage{booktabs}
\usepackage{mathtools}
\usepackage{multirow}
\usepackage{makecell}
\allowdisplaybreaks

\begin{document}
\title{Infrared divergences in the $\mathcal{O}(\alpha_s v^2)$ corrections to exclusive $J/\psi$ photoproduction}

\author{Zi-Qiang Chen}
\email{chenziqiang13@gzhu.edu.cn}
\affiliation{School of Physics and Materials Science, Guangzhou University, Guangzhou 510006, China}
\author{Long-Bin Chen}
\email{chenlb@gzhu.edu.cn}
\affiliation{School of Physics and Materials Science, Guangzhou University, Guangzhou 510006, China}


\begin{abstract}
We investigate the infrared structure of exclusive $J/\psi$ photoproduction, $\gamma p\to J/\psi p$, up to relative order $\alpha_s v^2$ within the frameworks of generalized parton distribution (GPD) and nonrelativistic QCD (NRQCD) factorization. 
By analyzing the singular loop-momentum regions, we extract the soft and collinear singular terms and identify residual infrared poles at relative order $\alpha_s v^2$ after QCD renormalization, collinear factorization into the GPDs, and NRQCD matching.
These poles originate from incomplete cancellation of soft-gluon exchange between the heavy-quark pair and the target-collinear partons.
We verify the pole coefficients through a complete one-loop calculation. 
We examine the local NRQCD operator basis and find no additional independent vacuum-to-$J/\psi$ matrix element that can absorb these poles.
These findings motivate further investigation of factorization in exclusive $J/\psi$ photoproduction.
\end{abstract}

\maketitle

\newpage
\section{Introduction}
Exclusive $J/\psi$ photoproduction, $\gamma p\to J/\psi p$, has attracted considerable interest.
The process probes gluon generalized parton distributions (GPDs), providing access to the three-dimensional gluon structure of the proton \cite{Ryskin:1992ui,Ji:1996ek,Hoodbhoy:1996zg,Radyushkin:1997ki,Burkardt:2000za,Ji:2003ak}.
Near threshold, it also offers a means of constraining gluonic gravitational form factors \cite{Mamo:2019mka,Guo:2021ibg,Guo:2023qgu,Guo:2025jiz}.
These opportunities have motivated extensive measurements at HERA \cite{ZEUS:1997kpo,ZEUS:2002wfj,H1:2005dtp,H1:2013okq}, the LHC \cite{ALICE:2014eof,LHCb:2018rcm}, and Jefferson Lab \cite{GlueX:2019mkq,GlueX:2023pev,Duran:2022xag}, with further prospects at the Electron-Ion Collider \cite{Boer:2024ylx}.
The process is also well suited to a theoretical treatment: the charm-quark mass provides a hard scale even for real photon, motivating a QCD factorization description that separates contributions associated with different scales.
Extracting proton structure from such measurements requires a controlled description of the production amplitude, including perturbative and relativistic corrections.

Within the GPD framework, exclusive $J/\psi$ photoproduction can be described in the kinematic region
\begin{equation}
\Lambda_{\mathrm{QCD}},\; m_N,\; \sqrt{-t} \ll m_{J/\psi},\; W,
\label{eq_region}
\end{equation}
where $m_N$ and $m_{J/\psi}$ are the proton and meson masses, respectively, $W$ is the photon--proton center-of-mass energy, and $t$ is the squared four-momentum transferred to the proton.
At leading twist, the amplitude is assumed to take the factorized form
\begin{equation}
\mathcal{M}^{\gamma p\to J/\psi p}=\sum_{\mathcal Q=g,q}\int_{-1}^{1}\mathrm{d}x\,T^{\gamma\mathcal Q\to J/\psi\mathcal Q}(x,\xi)F^{\mathcal Q}(x,\xi,t).
\label{eq_fac}
\end{equation}
Here $x$ is the longitudinal parton momentum fraction and $\xi$ is the skewness.
The proton matrix elements $F^{\mathcal Q}$ encode the unpolarized quark and gluon GPDs, whose definitions and parameterizations can be found in the reviews \cite{Diehl:2003ny,Belitsky:2005qn,Boer:2025ixc}.
Within the nonrelativistic QCD (NRQCD) factorization framework~\cite{Bodwin:1994jh}, the production amplitudes $T^{\gamma\mathcal Q\to J/\psi\mathcal Q}$ can be further expressed in terms of perturbatively calculable short-distance coefficients and vacuum-to-$J/\psi$ NRQCD matrix elements.
The latter encode the nonperturbative formation of the quarkonium.
This framework organizes the production amplitude as a double expansion in the strong coupling constant $\alpha_s$ and the heavy-quark relative velocity $v$.

Within this combined GPD--NRQCD framework, the next-to-leading order (NLO) QCD corrections to exclusive $J/\psi$ photoproduction have been calculated, confirming factorization at one loop and at leading order in the velocity expansion \cite{Ivanov:2004vd}.
That calculation includes both gluon and light-quark coefficient functions, with the light-quark contribution first appearing at NLO.
The corresponding NLO QCD corrections have also been worked out for exclusive electroproduction, $\gamma^*p\to J/\psi p$, with a spacelike photons of virtuality $Q^2>0$ \cite{Chen:2019uit,Flett:2021ghh}.
Subsequent developments include prescriptions to improve perturbative stability through appropriate factorization scale choices and low-virtuality subtractions \cite{Jones:2015nna, Jones:2016ldq, Flett:2019pux}, as well as the resummation of high-energy logarithms \cite{Flett:2024htj}.
The universal moment parameterization approach has been applied to extract small-$x$ gluon GPDs by combining exclusive $J/\psi$ production data with gluon PDF constraints \cite{Guo:2022upw,Guo:2023ahv,Guo:2024wxy}.
The $\mathcal{O}(v^2)$ correction to the tree-level photoproduction amplitude has been computed, revealing sizable effects for $J/\psi$ near threshold \cite{Blask:2025jua}.
The resulting double poles at $x=\pm\xi$ impose stronger smoothness requirements on the GPDs for a well-defined convolution.
Relativistic corrections have been incorporated into the high-energy-resummed coefficient functions \cite{Nefedov:2026cgj}.
A complete fixed-order calculation of the mixed $\mathcal{O}(\alpha_s v^2)$ correction, however, has not yet been reported.

In this work, we investigate the infrared structure of the mixed $\mathcal{O}(\alpha_s v^2)$ correction and examine whether the resulting singularities can be absorbed into the standard GPDs and NRQCD matrix elements.
By analyzing the relevant loop-momentum regions in both the gluon and light-quark channels, we identify infrared poles that remain after QCD renormalization, collinear factorization into the GPDs, and NRQCD matching.
We cross-check the region analysis against a complete one-loop calculation.
The residual poles arise because soft-gluon exchanges between the heavy-quark pair and the partons collinear to the proton cancel only at leading order in $v$.
Similar infrared divergences were discussed for exclusive double-quarkonium production in Ref.~\cite{Bodwin:2010fi}.
There, the corresponding contributions are suppressed by powers of the heavy-quark mass over the collision energy, whereas the residual poles found here exhibit no analogous suppression by powers of $m_{J/\psi}/W$.
How the factorized form in Eq.~(\ref{eq_fac}) should be extended beyond leading order in $v$ therefore remains open.

The rest of the paper is organized as follows. 
In Section II, we introduce the NRQCD factorization formula and present the leading-order results. 
In Section III, we analyze the infrared structure of the light-quark channel and discuss the origin of the residual divergences. 
In Section IV, we present the corresponding results for the gluon channel.
The final section presents our summary and discussion.

\section{NRQCD factorization formula and leading-order results}
Within the NRQCD factorization framework \cite{Bodwin:1994jh,Bodwin:2010fi}, the production amplitude $T^{\gamma\mathcal Q\to J/\psi\mathcal Q}$ up to order $v^2$ can be expressed as
\begin{equation}
T^{\gamma\mathcal Q\to J/\psi\mathcal Q}=\sqrt{2m_{J/\psi}}\left[c_0\langle J/\psi|\psi^\dagger\boldsymbol{\sigma}\cdot\boldsymbol{\epsilon}\chi|0\rangle+\frac{c_2}{m_c^2}\langle J/\psi|\psi^\dagger\left(-\tfrac{i}{2}\overleftrightarrow{\boldsymbol D}\right)^2\boldsymbol{\sigma}\cdot\boldsymbol{\epsilon}\chi|0\rangle\right],
\label{eq_amp1}
\end{equation}
where $c_0$ and $c_2$ are short-distance coefficients, $m_c$ is the charm-quark mass, $\langle J/\psi|\psi^\dagger\boldsymbol{\sigma}\cdot\boldsymbol{\epsilon}\chi|0\rangle$ and $\langle J/\psi|\psi^\dagger\left(-\tfrac{i}{2}\overleftrightarrow{\boldsymbol D}\right)^2\boldsymbol{\sigma}\cdot\boldsymbol{\epsilon}\chi|0\rangle$ are the vacuum-to-$J/\psi$ NRQCD matrix elements.
The short-distance coefficients $c_0$ and $c_2$ are determined through the standard matching procedure \cite{Bodwin:1994jh}, in which the physical $J/\psi$ state is replaced by a free $c\bar c$ pair with quantum numbers ${}^3S_1^{[1]}$. 
The matching condition reads
\begin{align}
T^{\gamma\mathcal Q\to c\bar c({}^3S_1^{[1]})\mathcal Q}\Big|_{\mathrm{pert\,QCD}}={}&c_0\langle c\bar c({}^3S_1^{[1]})|\psi^\dagger\boldsymbol{\sigma}\cdot\boldsymbol{\epsilon}\chi|0\rangle\Big|_{\mathrm{pert\,NRQCD}}
\nonumber\\
&+\frac{c_2}{m_c^2}\langle c\bar c({}^3S_1^{[1]})|\psi^\dagger\left(-\tfrac{i}{2}\overleftrightarrow{\boldsymbol D}\right)^2\boldsymbol{\sigma}\cdot\boldsymbol{\epsilon}\chi|0\rangle\Big|_{\mathrm{pert\,NRQCD}}.
\label{eq_match}
\end{align}
Here, the perturbative NRQCD matrix elements can themselves contain infrared poles, which must be included in the matching.

To evaluate the QCD amplitude in Eq.~(\ref{eq_match}), we consider the partonic process
\begin{equation}
\gamma(k_\gamma)+\mathcal Q(k_1) \to c\left(\frac{K}{2}+q\right)\bar c\left(\frac{K}{2}-q\right)+\mathcal Q(k_2),
\end{equation}
where $K$ and $q$ denote the total and relative momenta of the heavy-quark pair, respectively.
The external partons satisfy the on-shell conditions $k_\gamma^2=k_1^2=k_2^2=0$, $(\frac K2+q)^2=(\frac K2-q)^2=m_c^2$.
We work in the $c\bar c$ rest frame and consider forward scattering, corresponding to the minimum kinematically allowed value of $|t|$.
Neglecting the proton mass in the hard-subprocess kinematics, we parameterize the momenta as
\begin{align}
&k_\gamma=En_-,\qquad k_1=\frac{E}{2}\frac{x+\xi}{\xi}n_+,\qquad k_2=\frac{E}{2}\frac{x-\xi}{\xi}n_+,\nonumber\\
&K=E(n_++n_-),\qquad q=\frac{q_z}{2}(n_+-n_-)+q_\perp,
\end{align}
where $n_\pm=(1,\boldsymbol 0_\perp,\pm1)$ are lightlike reference vectors and $E=\sqrt{m_c^2+\boldsymbol q^2}$.
Here and throughout this paper, boldface symbols denote spatial vectors.
In dimensional regularization with $D=4-2\epsilon$, the transverse vectors belong to a $(D-2)$-dimensional space, with $\boldsymbol q^2=\boldsymbol q_\perp^2+q_z^2$.
In this kinematics, the skewness is related to the photon–proton center-of-mass energy by $\xi=2E^2/(W^2-2E^2)$.
Since the amplitude is ultimately proportional to $\varepsilon_\gamma\cdot\varepsilon_{J/\psi}^{*}$, we choose, for convenience, a common real transverse polarization, $\varepsilon_{J/\psi}=\varepsilon_\gamma=\varepsilon_\perp$, with $\varepsilon_\perp^2=-1$ and $\varepsilon_\perp\cdot n_\pm=0$.

We project the heavy-quark pair into the spin-triplet, color-singlet state using the covariant projector \cite{Bodwin:2002cfe}
\begin{equation}
\Pi_{c\bar{c}}=\frac{1}{4\sqrt{2}E(E+m_c)}\left(\frac{\slashed K}{2}-\slashed q-m_c\right)\slashed{\varepsilon}_{J/\psi}^{*}(\slashed K+2E)\left(\frac{\slashed K}{2}+\slashed q+m_c\right)\otimes\frac{\mathbf 1_c}{\sqrt{N_c}},
\end{equation}
where $\mathbf 1_c$ is the identity matrix in the fundamental color representation and $N_c=3$.
For the unpolarized GPD sector, we use the standard spin and color projectors \cite{Belitsky:2005qn}:
\begin{align}
&\Pi_{gg}^{\mu\nu}=-\frac{g_\perp^{\mu\nu}}{2(D-2)(x^2-\xi^2)}\otimes\frac{\mathbf1_A}{N_c^2-1},\quad \text{for the gluon channel},\\
&\Pi_{q\bar{q}}=\frac{E}{4\xi}\slashed n_+\otimes\frac{\mathbf1_c}{N_c},\quad \text{for the light-quark channel}.
\end{align}
Here $\mathbf1_A$ is the identity matrix in the adjoint color representation and $g_\perp^{\mu\nu} =g^{\mu\nu}-\tfrac12(n_+^\mu n_-^\nu+n_-^\mu n_+^\nu)$.
The gluon projector includes an average over the $D-2$ transverse polarizations and a symmetry factor of $1/2$.

\begin{figure}
\centering
\includegraphics[scale=0.55]{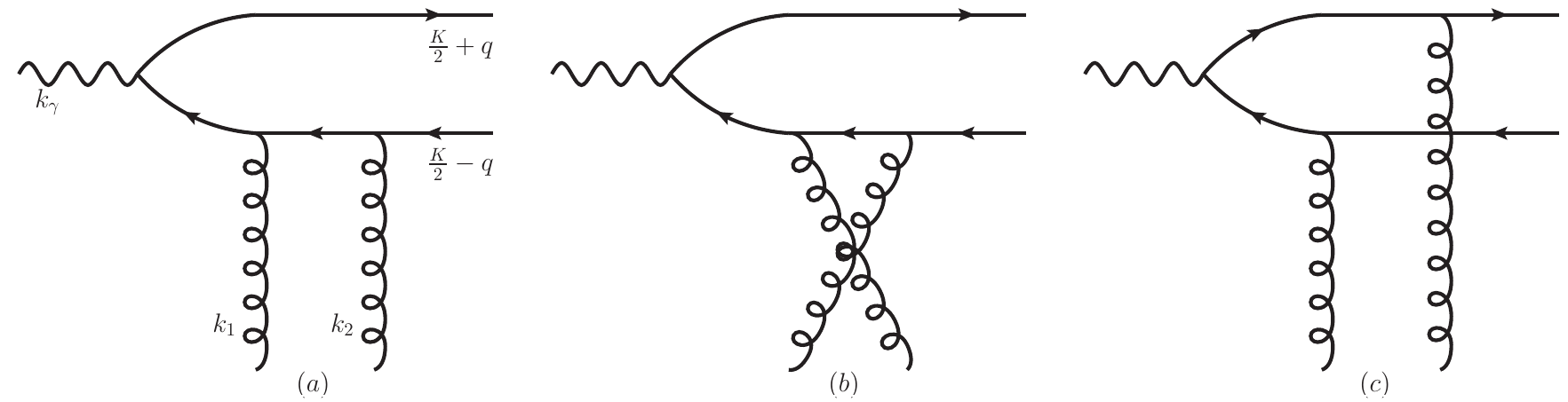}
\caption{Three of the six LO Feynman diagrams. The remaining three are obtained by reversing the heavy-quark line.}
\label{fig_feynLO}
\end{figure}

At leading order (LO) in $\alpha_s$, only the gluon channel contributes. 
There are six Feynman diagrams, three of which are shown in Fig.~\ref{fig_feynLO}. 
Keeping the full relative-momentum dependence, we write the spin-triplet, color-singlet amplitude before the $S$-wave average as
\begin{align}
T_{\mathrm{LO}}(x,\xi;q)=&-\frac{2}{3}\sqrt{\frac{2}{3}}\,\frac{eg_s^2}{(D-2)(E^2-q_z^2)(x^2-\xi^2)}\Bigg\{E\left[D-2+\frac{4\xi^2\boldsymbol q_\perp^2}{(E^2-q_z^2)(x^2-\xi^2)}\right] \nonumber\\
& -\frac{(q\cdot\varepsilon_\perp)^2}{E+m_c}\left[D-2-\frac{4\xi^2\left[m_c(E+m_c)-\boldsymbol q_\perp^2\right]}{(E^2-q_z^2)(x^2-\xi^2)}\right]\Bigg\}.
\label{eq_LO_unaveraged}
\end{align}
Averaging over the direction of $\boldsymbol q$ at fixed $\boldsymbol q^2$, we obtain
\begin{align}
\left\langle T_{\mathrm{LO}}(x,\xi;q)\right\rangle_S=-\frac{8}{3}\sqrt{\frac{2}{3}}\,\frac{eg_s^2}{(D-2)E(x^2-\xi^2)}\left[\frac{\mathcal F_D}{4}+\frac{\xi^2\mathcal G_D}{x^2-\xi^2}\right],
\label{eq_LO_averaged}
\end{align}
where
\begin{align}
\mathcal F_D={}&\left(D-2+\frac{m_c^2}{E(E+m_c)}\right){}_2F_1\!\left(1,\frac12;\frac{D-1}{2};\frac{\boldsymbol q^2}{E^2}\right)-\frac{E}{E+m_c},\\
\mathcal G_D={}&\frac{1}{2(D-2)}\Bigg\{\left[\left(D-3+\frac{2m_c}{E}\right)\left(D-2-\frac{m_c^2}{E^2}\right)+\frac{(D-2)E^2+m_c^2}{E(E+m_c)}\right]\nonumber\\
&\times{}_2F_1\!\left(1,\frac12;\frac{D-1}{2};\frac{\boldsymbol q^2}{E^2}\right)-(D-3)\left(D-3+\frac{2m_c}{E}\right)-\frac{(D-1)E}{E+m_c}\Bigg\}.
\end{align}
Here ${}_2F_1$ denotes the Gauss hypergeometric function. 
Setting $D=4$, we recover the result given in Ref.~\cite{Nefedov:2026cgj}.

For the relativistic expansion, we choose to hold $E$ fixed. 
This keeps the skewness $\xi$ entering the GPDs unchanged at a given $W$. 
After perturbative matching, the physical $J/\psi$ amplitude can be expressed in terms of the measured meson mass by identifying $2E$ with $m_{J/\psi}$ to the accuracy considered. Substituting $m_c=\sqrt{E^2-\boldsymbol q^2}$ into Eq.~(\ref{eq_LO_averaged}) and expanding up to $\mathcal O(v^2)$, we obtain
\begin{align}
\left\langle T_{\mathrm{LO}}(x,\xi;q)\right\rangle_S={}&-\frac{2}{3}\sqrt{\frac{2}{3}}\,\frac{eg_s^2}{E(x^2-\xi^2)}\Bigg[1+\frac{\boldsymbol q^2\left[(D-2)x^2+(7D-6)\xi^2\right]}{2(D-1)(D-2)E^2(x^2-\xi^2)}+\mathcal O(v^4)\Bigg].
\label{eq_LO_v2}
\end{align}

\section{Infrared structure of the light-quark channel at NLO}
In this section, we analyze the infrared structure of the light-quark channel. 
We examine the singular loop-momentum regions and use power counting to extract the corresponding singular terms. 
Since the amplitude in this channel first appears at one loop, no QCD ultraviolet counterterms or contributions from one-loop NRQCD matrix elements are required at this order. 
This makes the light-quark channel a relatively simple starting point for examining the cancellation of singularities. 
The analysis must still include a nonzero counterterm from GPD renormalization, which involves the gluon Born amplitude through quark–gluon mixing.

\subsection{Collinear singularities and GPD renormalization}
\begin{figure}
\centering
\includegraphics[scale=0.55]{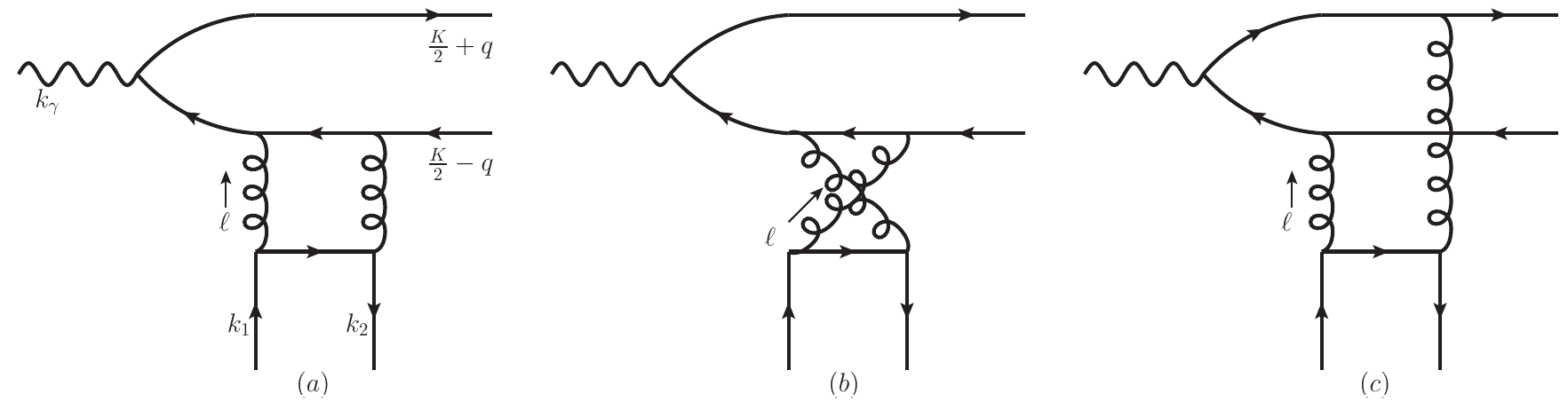}
\caption{Three of the six one-loop Feynman diagrams for the light-quark channel. The remaining three are obtained by reversing the heavy-quark line.}
\label{fig_feynNLO_q}
\end{figure}

There are six one-loop Feynman diagrams contributing to the light-quark channel, three of which are shown in Fig.~\ref{fig_feynNLO_q}. 
With the loop-momentum assignments shown in the figure, the region collinear to the proton direction can be parameterized as 
\begin{equation}
\ell=Ez\,n_++\lambda\ell_T+\lambda^2\beta\,n_-,
\label{eq_coll_scaling}
\end{equation}
where $n_\pm\cdot\ell_T=0$, and $z$, $\beta$, and $\ell_T$ are held fixed as $\lambda\to0$. 
The corresponding integration measure is $\mathrm d^D\ell=E\lambda^D\,\mathrm dz\,\mathrm d\beta\,\mathrm d(\boldsymbol\ell_T^{\,2})\,\mathrm d\Omega_\perp\,(\boldsymbol\ell_T^{\,2})^{-\epsilon}$, where $\mathrm d\Omega_\perp$ denotes the solid-angle element in the $(D-2)$-dimensional transverse space.

With these assignments, all six diagrams share the same three massless loop propagators, whose denominators are
\begin{align}
D_1&=\ell^2+i0=(4Ez\beta-\boldsymbol\ell_T^{\,2})\lambda^2+i0,\nonumber\\
D_2&=(\ell-k_1)^2+i0=[2E\beta(2z-1-r)-\boldsymbol\ell_T^{\,2}]\lambda^2+i0,\nonumber\\
D_3&=(\ell-k_1+k_2)^2+i0=[4E\beta(z-1)-\boldsymbol\ell_T^{\,2}]\lambda^2+i0,
\end{align}
where $r=x/\xi$. 
In the following analysis, we work in the ERBL region, $-\xi<x<\xi$, with $\xi>0$, so that $-1<r<1$.
Away from the soft endpoints, the heavy-quark propagators remain hard in the collinear limit and are included in the hard coefficient.
After summing the contributions of all six diagrams and performing the transverse angular integration, we obtain
\begin{align}
T_{\mathrm{NLO,coll}}^q={}&\frac{4i g_s^2 E\xi}{3\pi^3}\frac{(4\pi\mu_R^2)^\epsilon}{\Gamma(1-\epsilon)}\int_{\mathrm{coll}}\mathrm dz\,\mathrm d\beta\,\mathrm d(\boldsymbol\ell_T^{\,2})\,(\boldsymbol\ell_T^{\,2})^{-\epsilon}\nonumber\\
&\times T_{\mathrm{LO}}(\xi(2z-1),\xi;q)\frac{4E\beta z(1-z)+\frac{2r}{D-2}\boldsymbol\ell_T^{\,2}}{d_1d_2d_3}.
\end{align}
Here, $d_i=D_i/\lambda^2$, and the Born amplitude $T_{\mathrm{LO}}$ is evaluated at $x=\xi(2z-1)$.

Performing the $\beta$ integration by closing the contour in either the upper or lower half-plane gives a nonzero contribution only for $0<z<1$. 
Since the pole configurations differ between $0<z<c$ and $c<z<1$, where $c=(1+r)/2\in(0,1)$, the two intervals are treated separately. 
The remaining integration over $\boldsymbol\ell_T^{\,2}$ is logarithmically divergent and generates the collinear pole:
\begin{align}
T_{\mathrm{NLO,coll}}^q=-\frac{2g_s^2\xi}{3\pi^2\epsilon}\int_{\delta_0}^{1-\delta_1}\mathrm dz\,\left[\mathcal K(z,r;D)\,T_{\mathrm{LO}}(\xi(2z-1),\xi;q)\right]_{D=4},
\label{eq_coll_pole}
\end{align}
Here, we introduce two independent cutoffs, $\delta_0$ and $\delta_1$, to exclude the soft endpoint regions from the collinear contribution. 
These cutoffs match the soft regions with spatial radii $E\delta_0$ and $E\delta_1$, respectively.
The kernel $\mathcal K(z,r;D)$ in Eq.~(\ref{eq_coll_pole}) takes the form
\begin{equation}
\mathcal K(z,r;D)=
\begin{cases}
\displaystyle
\frac{z}{c}\left(1-z+\frac{2r}{D-2}\right),&0<z<c,
\\[8pt]
\displaystyle
\frac{z-1}{1-c}\left(z-\frac{2r}{D-2}\right),&c<z<1.
\end{cases}
\label{eq_coll_weight}
\end{equation}
The value of $\mathcal K$ at the isolated point $z=c$ does not affect the integration over $z$.

The remaining integral over $z$ can be evaluated analytically, yielding
\begin{align}
T_{\mathrm{NLO,coll}}^q={}&\frac{2}{9}\sqrt{\frac23}\,\frac{e g_s^4}{\pi^2\xi\epsilon}\Bigg\{\frac{r}{r^2-1}\,\frac{E-\frac{(q\cdot\varepsilon_\perp)^2}{E+m_c}}{E^2-q_z^2}\left[
\ln\frac{4}{1-r^2}-2r\operatorname{arctanh}r\right]  \nonumber\\
&+\frac{E\boldsymbol q_\perp^{\,2}+\frac{(q\cdot\varepsilon_\perp)^2}{E+m_c}\left[m_c(E+m_c)-\boldsymbol q_\perp^{\,2}\right]}{(E^2-q_z^2)^2}\left[2\operatorname{arctanh}r-\frac{r}{r^2-1}-\frac12\ln\frac{\delta_0}{\delta_1}\right]\Bigg\}.
\label{eq_coll_pole_integrated}
\end{align}
Here, we have neglected terms of order $\delta_0$ or $\delta_1$.

The collinear counterterm arises from the renormalization of the GPDs in the $\overline{\mathrm{MS}}$ scheme. 
Since the light-quark channel has no Born contribution, this counterterm is given by the convolution of the gluon Born amplitude with the leading-order evolution kernel $V_{gq}^{(0)}$ that mixes the quark GPD into the gluon GPD:
\begin{equation}
T_{\mathrm{NLO,GPD}}^q=\frac{\alpha_s}{2\pi\epsilon}\int_{-1}^{1}\mathrm dy\,\left.T_{\mathrm{LO}}(y,\xi;q)\right|_{D=4}V_{gq}^{(0)}(y,x;\xi).
\label{eq_gpd_counterterm}
\end{equation}
This counterterm cancels the cutoff-independent terms in Eq.~(\ref{eq_coll_pole_integrated}), leaving
\begin{align}
T_{\mathrm{NLO,coll}}^q+T_{\mathrm{NLO,GPD}}^q={}&-\frac{1}{9}\sqrt{\frac23}\,\frac{e g_s^4}{\pi^2\xi\epsilon}\frac{E\boldsymbol q_\perp^{\,2}+\frac{(q\cdot\varepsilon_\perp)^2}{E+m_c}\left[m_c(E+m_c)-\boldsymbol q_\perp^{\,2}\right]}{(E^2-q_z^2)^2}\ln\frac{\delta_0}{\delta_1}.
\label{eq_gpd_coll_cancellation}
\end{align}
For equal cutoffs, $\delta_0=\delta_1$, the collinear pole is completely canceled by the GPD counterterm.

\subsection{Soft singularities}
We next consider the soft regions around $\ell=0$ and $\ell=En_+$, which were excluded from the collinear contribution. To extract their singular terms, we use the parameterizations $\ell=\lambda h$ and $\ell=En_+-\lambda h$, respectively, and take $\lambda\to0$ with all components of $h$ held fixed. Expanding the integrands in $\lambda$ and summing the contributions of all six diagrams, we obtain
\begin{align}
T_{\mathrm{NLO,soft}}^q={}&-\frac{i e g_s^4\mu_R^{2\epsilon}}{18\sqrt6\,E^2(E+m_c)\xi}\Bigg[\frac{1}{1+r}\int_{|\boldsymbol h|<\delta_0E}\frac{\mathrm d^Dh}{(2\pi)^D}\frac{r(J\cdot n_+)(H\cdot h)+(n_+\cdot h)(J\cdot H)}{(h^2+i0)(n_+\cdot h-i0)^2}\nonumber\\
&+\frac{1}{1-r}\int_{|\boldsymbol h|<\delta_1E}\frac{\mathrm d^Dh}{(2\pi)^D}\frac{r(J\cdot n_+)(H\cdot h)-(n_+\cdot h)(J\cdot H)}{(h^2+i0)(n_+\cdot h-i0)^2}\Bigg].
\label{eq_soft_integrand}
\end{align}
Here, $J^\mu$ is the heavy-quark dipole current, and $H^\mu$ is the hard current for $\gamma g_{\rm coll}\to c\bar c$, with $g_{\rm coll}$ denoting the gluon collinear to the proton.
With the overall color factor included in the prefactor of Eq.~(\ref{eq_soft_integrand}), these two currents take the form
\begin{align}
\label{eq_soft_dipole}
&J^\mu(h,q)=\frac{k_c^\mu}{k_c\cdot h-i0}-\frac{k_{\bar c}^\mu}{k_{\bar c}\cdot h-i0}, \\
\label{eq_soft_hard_current}
&H^\mu(q)=\frac{16E}{E^2-q_z^2}\bigg\{-\left[E(E+m_c)-2(q\cdot\varepsilon_\perp)^2\right]q_\perp^\mu+(q\cdot\varepsilon_\perp)^2q_z n_+^\mu\nonumber \\
&\qquad\qquad+\left[E(E+m_c)-q_z^2\right](q\cdot\varepsilon_\perp)\varepsilon_\perp^\mu\bigg\},
\end{align}
where $k_c=K/2+q$ and $k_{\bar c}=K/2-q$. 
Together, these two currents describe soft-gluon exchange between the light-quark line and the heavy-quark pair.

We perform the $h^0$ integration by closing the contour and evaluate the remaining spatial integral in $D-1$ dimensions. 
The resulting soft pole is
\begin{align}
T_{\mathrm{NLO,soft}}^q={}&\frac{e g_s^4}{72\sqrt6\,\pi^2(E+m_c)E^2}\frac{1}{\epsilon}\left[\frac{x}{x^2-\xi^2}+\frac{1}{2\xi}\ln\frac{\delta_0}{\delta_1}\right]\left[\frac{k_c\cdot H}{k_c\cdot n_+}-\frac{k_{\bar c}\cdot H}{k_{\bar c}\cdot n_+}\right].
\label{eq_soft_pole_current}
\end{align}
Substituting the explicit expression for $H^\mu$, we obtain
\begin{equation}
T_{\mathrm{NLO,soft}}^q=\frac{2}{9}\sqrt{\frac23}\,\frac{e g_s^4}{\pi^2\epsilon}\left[\frac{x}{x^2-\xi^2}+\frac{1}{2\xi}\ln\frac{\delta_0}{\delta_1}\right]\frac{E\boldsymbol q_\perp^{\,2}+\frac{(q\cdot\varepsilon_\perp)^2}{E+m_c}\left[m_c(E+m_c)-\boldsymbol q_\perp^{\,2}\right]}{(E^2-q_z^2)^2}.
\label{eq_soft_pole_unaveraged}
\end{equation}
The cutoff-dependent term cancels against the contribution given in Eq.~(\ref{eq_gpd_coll_cancellation}), leaving a pole independent of the cutoff parameters.
This result relies on the soft approximation used in Eq.~(\ref{eq_soft_integrand}).
One subregion where that approximation could fail is the Glauber region, which we examine next.

\begin{figure}
\centering
\includegraphics[width=\textwidth]{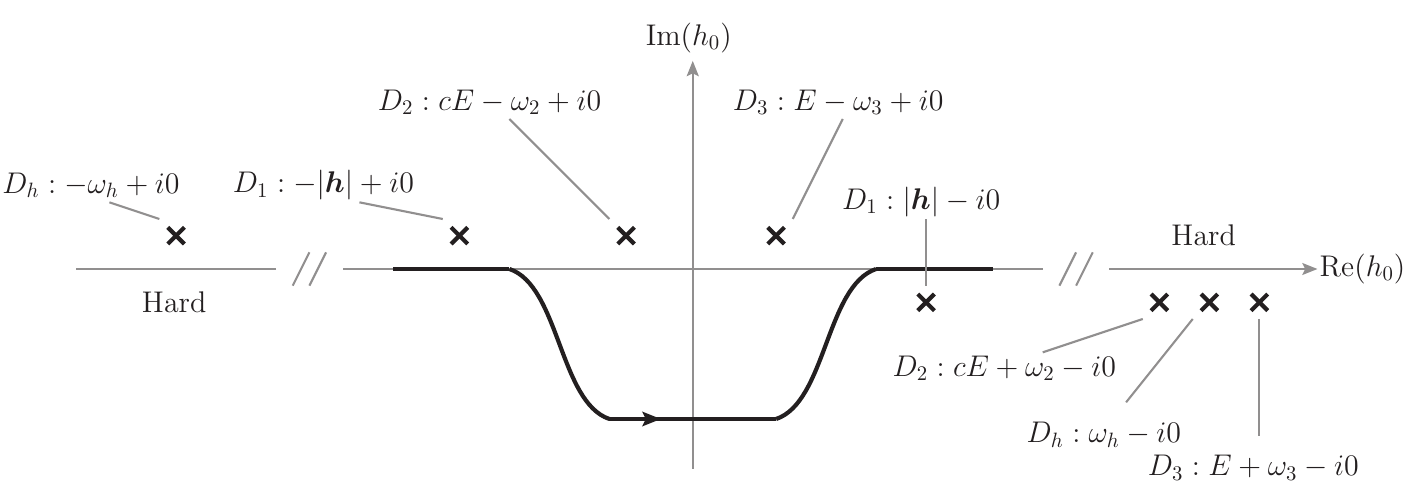}
\caption{Poles and contour deformation in the complex $h^0$ plane for Fig.~\ref{fig_feynNLO_q}(a) near $\ell=0$. Here, $\omega_2=|cE\hat{\boldsymbol z}-\boldsymbol h|$, $\omega_3=|E\hat{\boldsymbol z}-\boldsymbol h|$, and $\omega_h=\sqrt{m_c^2+|E\hat{\boldsymbol z}+\boldsymbol q-\boldsymbol h|^2}$, with $\hat{\boldsymbol z}$ the spatial unit vector.}
\label{fig_soft_pole}
\end{figure}

In the Glauber region, $|h^0|$ and $|h_z|$ can be much smaller than $|\boldsymbol h_\perp|$, so that $h^2$ need not be suppressed relative to the linear terms in the near-on-shell propagators. 
We illustrate the argument with diagram (a) in Fig.~\ref{fig_feynNLO_q} near the endpoint $\ell=0$.
The exact loop propagator denominators are
\begin{align}
D_1&=h^2+i0,\nonumber\\
D_2&=(h-cEn_+)^2+i0,\nonumber\\
D_3&=(h-En_+)^2+i0,\nonumber\\
D_h&=(h-\tfrac{E}{2}n_++\tfrac{E}{2}n_--q)^2-m_c^2+i0.
\label{eq_glaA_denominators}
\end{align}
We assume $|h^0|,|\boldsymbol h|\ll cE,E$, without imposing a hierarchy among the components of $h$. 
The pole configuration of these four denominators in the complex $h^0$ plane is illustrated in Fig.~\ref{fig_soft_pole}. 
Of the eight poles, four lie in the soft region and four remain at hard energies. 
The original integration over $h^0$ runs along the real axis, passing through the small-energy region near the origin where Glauber configurations can occur.
To avoid these configurations, we deform the central segment into the lower half-plane, without crossing a pole, as shown in Fig.~\ref{fig_soft_pole}.
The deformation is chosen to give the central segment an imaginary part of order $|\boldsymbol h|$, while reconnecting to the real axis before reaching the lower gluon pole. 
For a near-on-shell denominator $h^2-2P\cdot h+i0$, with $P=cEn_+$ or $En_+$ in this diagram, the central deformed segment satisfies
\begin{equation}
|P\cdot h|\sim P^0|\boldsymbol h|,\qquad |h^2|=\mathcal O(|\boldsymbol h|^2),\qquad \left|\frac{h^2}{2P\cdot h}\right|=\mathcal O\!\left(\frac{|\boldsymbol h|}{P^0}\right),
\label{eq_glauber_bound}
\end{equation}
which justifies the eikonal expansion of the near-on-shell propagators.
We have also checked the numerator expansions and verified that the omitted terms do not contribute to infrared poles on the deformed contour.
A similar argument holds for the other diagrams and for the endpoint $\ell=En_+-h$. 
The Glauber region therefore does not modify the soft poles obtained above.

\subsection{Origin and implications of the residual infrared pole}
Combining the collinear contribution, the GPD counterterm, and the soft contribution, we obtain the residual infrared pole
\begin{align}
T_{\mathrm{NLO,residual}}^q={}&\frac{2}{9}\sqrt{\frac23}\,\frac{e g_s^4}{\pi^2\epsilon}\frac{x}{x^2-\xi^2}\frac{E\boldsymbol q_\perp^{\,2}+\frac{(q\cdot\varepsilon_\perp)^2}{E+m_c}\left[m_c(E+m_c)-\boldsymbol q_\perp^{\,2}\right]}{(E^2-q_z^2)^2}.
\label{eq_quark_residual_pole_unaveraged}
\end{align}
After averaging over the direction of $\boldsymbol q$ and expanding up to $\mathcal O(v^2)$ at fixed $E$, we obtain
\begin{equation}
\left\langle T_{\mathrm{NLO,residual}}^q\right\rangle_S=\frac{2}{9}\sqrt{\frac23}\,\frac{e g_s^4}{\pi^2E}\frac{x}{x^2-\xi^2}\frac{1}{\epsilon}\left[\frac{\boldsymbol q^{\,2}}{E^2}+\mathcal O(v^4)\right].
\label{eq_soft_pole_v2}
\end{equation}
We have verified this result through a complete one-loop calculation. 
Although the analysis above is carried out in the ERBL region, the residual pole remains nonzero upon analytic continuation to the DGLAP region.

\begin{figure}
\centering
\includegraphics[scale=0.55]{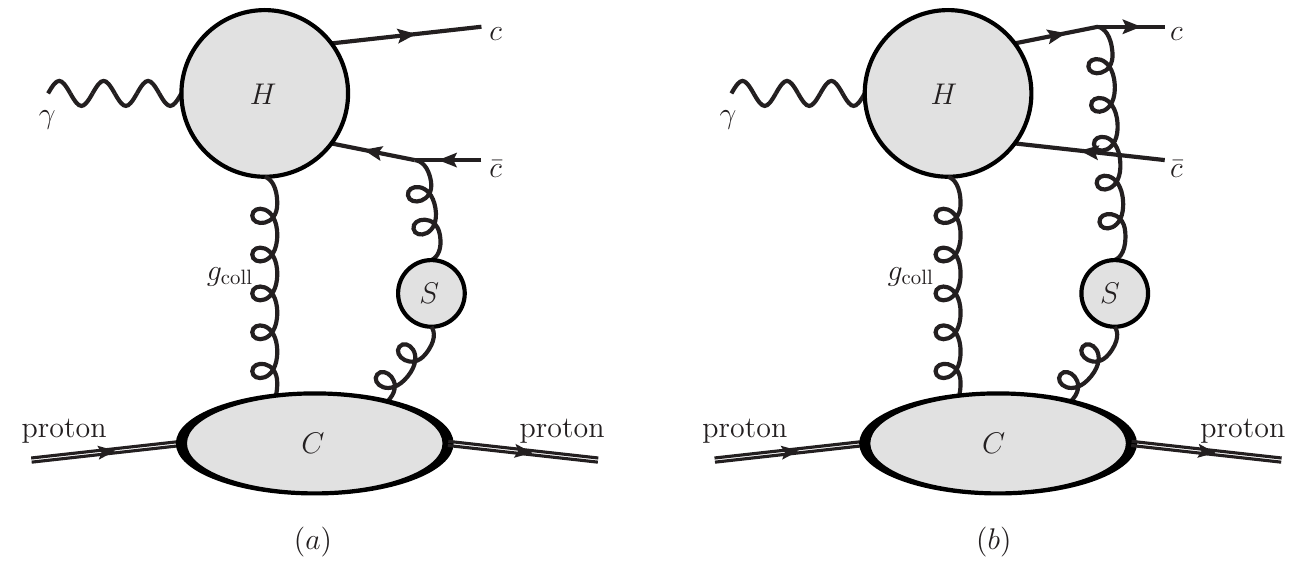}
\caption{Reduced diagrams illustrating soft-gluon exchange between the target-collinear subgraph and the heavy antiquark (a) or quark (b).}
\label{fig_reduced}
\end{figure}

For equal endpoint cutoffs, $\delta_0=\delta_1$, the collinear pole is completely canceled by the GPD counterterm, so the residual pole comes entirely from the soft contribution. 
Its origin is apparent from the current structure in Eq.~(\ref{eq_soft_pole_current}), which describes soft-gluon exchange between the light-quark line and the heavy-quark pair. 
The quark and antiquark attachments enter with opposite color signs, as encoded in the dipole current $J^\mu(h,q)$. 
At $q=0$, the two heavy particles have the same velocity, so their eikonal factors coincide and $J^\mu(h,0)=0$. 
At nonzero $q$, this cancellation is incomplete, and the dipole current starts at linear order  in $q$. 
In the present channel, the hard current $H^\mu(q)$ also contains a term linear in $q$.
The product of these two linear terms produces the residual pole at order $v^2$.
The reduced diagrams in Fig.~\ref{fig_reduced} summarize the soft configurations discussed above.
Here, the hard subgraph $H$ corresponds to the current $H^\mu$ in Eq.~(\ref{eq_soft_hard_current}), and $C$ denotes the target-collinear subgraph.
For the one-loop diagrams in Fig.~\ref{fig_feynNLO_q}, $H$ represents the tree-level $\gamma g_{\rm coll}\to c\bar c$ subprocess, $C$ includes the target-collinear propagator and the $q\bar{q}g$ vertex, and $S$ reduces to a single soft-gluon propagator.

Such incomplete cancellation was also discussed for exclusive double-quarkonium production in Ref.~\cite{Bodwin:2010fi}. 
In both cases, soft gluons connect partonic sectors that are widely separated in rapidity: the two outgoing quarkonia in that process, and the target-collinear partons and the heavy-quark pair in the present one. 
However, the corresponding contributions in double-quarkonium production are suppressed by powers of $m_c/\sqrt{s}$, whereas the residual pole found here has no analogous suppression by powers of $m_{J/\psi}/W$.
In the reduced diagrams of Fig.~\ref{fig_reduced}, the hard subgraph $H$ is characterized by $(k_\gamma+k_{g_{\rm coll}})^2=4E^2\sim m_{J/\psi}^2$, which remains fixed as $W$ increases.

We next examine whether additional local NRQCD matrix elements associated with higher Fock components of the $J/\psi$ can absorb the residual pole.
The relevant local currents and their velocity scalings have been systematically discussed in Ref.~\cite{Brambilla:2019fmu}.
Within the basis and the velocity counting adopted there, no additional independent vacuum-to-$J/\psi$ matrix element contributes through relative order $v^2$ beyond those already included in Eq.~(\ref{eq_amp1}).
In particular, matrix elements containing explicit chromoelectric or chromomagnetic fields, such as $\langle J/\psi|\psi^\dagger g_s\boldsymbol B\cdot\boldsymbol\epsilon\chi|0\rangle$ and
$\langle J/\psi|\psi^\dagger\boldsymbol\sigma\cdot\boldsymbol\epsilon(\overleftrightarrow{\boldsymbol D}\cdot g_s\boldsymbol E+g_s\boldsymbol E\cdot\overleftrightarrow{\boldsymbol D})\chi|0\rangle$, contribute only at higher orders in $v$.
The residual pole therefore cannot be absorbed by simply adding a new local higher-Fock matrix element.

A useful comparison is provided by the leading-power factorization theorem proven for longitudinal light-meson electroproduction at large photon virtuality~\cite{Collins:1996fb}.
At leading power, the meson constituents share a common lightlike direction and hence identical eikonal factors, allowing their soft couplings to cancel in the color-singlet channel.
The same cancellation holds at $q=0$ for the heavy-quark pair, whose constituents share a common timelike velocity.
In the present photoproduction process, the relative motion of the heavy quarks distinguishes their eikonal factors, and the resulting soft contribution is not covered by that theorem.

\section{Infrared structure of the gluon channel at NLO}
We now turn to the gluon channel, whose infrared structure can be analyzed using the same method as for the light-quark channel.
The analysis is more involved because of the larger number of diagrams and the additional contributions from QCD renormalization and NRQCD matching.
We summarize the results below, focusing on the additional features of this channel.

\begin{figure}
\centering
\includegraphics[scale=0.55]{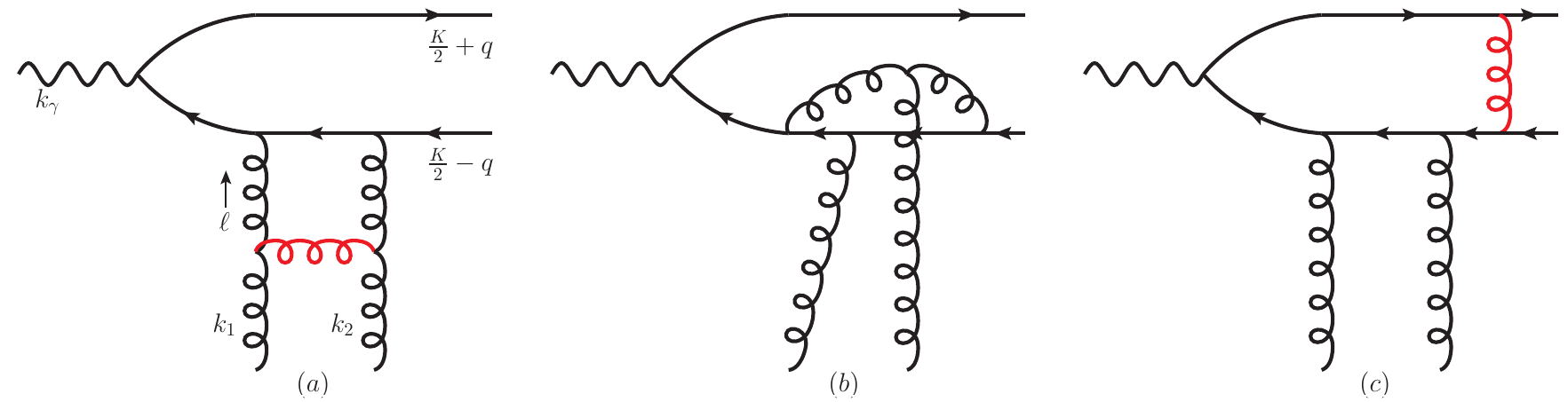}
\caption{Representative one-loop Feynman diagrams for the gluon channel.}
\label{fig_feynNLO_g}
\end{figure}

Representative one-loop Feynman diagrams for the gluon channel are shown in Fig.~\ref{fig_feynNLO_g}.
Since the gluon attachments are more varied than in the light-quark channel, we use a common spatial radius $\delta E$ around all soft centers to separate the soft and collinear contributions.
For example, the three massless loop propagators in Fig.~\ref{fig_feynNLO_g}(a) carry momenta $\ell$, $\ell-k_1$, and $\ell-k_1+k_2$.
Their soft neighborhoods are excluded by imposing $|\boldsymbol\ell|>\delta E$, $|\boldsymbol\ell-\boldsymbol k_1|>\delta E$, and $|\boldsymbol\ell-\boldsymbol k_1+\boldsymbol k_2|>\delta E$.
Summing the target-collinear contributions from all one-loop diagrams, we obtain the collinear pole of the bare amplitude:
\begin{align}
T_{\mathrm{NLO,coll}}^g=&\frac{e g_s^4}{2\sqrt6\,\pi^2\xi^2(1-r^2)^2\epsilon}\Bigg\{\frac{E-\frac{(q\cdot\varepsilon_\perp)^2}{E+m_c}}{E^2-q_z^2}\Bigg[r^2-1+(1+r^2)\left(\ln\frac{4}{1-r^2}-2r\operatorname{arctanh}r\right)\Bigg]\nonumber\\
&+\frac{E\boldsymbol q_\perp^{\,2}+\frac{(q\cdot\varepsilon_\perp)^2}{E+m_c}\left[m_c(E+m_c)-\boldsymbol q_\perp^{\,2}\right]}{(E^2-q_z^2)^2}\left[1-r^2+2r(r^2-3)\operatorname{arctanh}r\right]\Bigg\}.
\label{eq_gluon_coll_pole}
\end{align}
This result is obtained in the ERBL region, and the dependence on $\delta$ cancels in the sum because we use the same spatial cutoff around all soft centers.

The infrared contribution from on-shell renormalization of the external gluon fields is
\begin{equation}
T_{\mathrm{NLO},Z_3}^g=\frac{\alpha_s}{4\pi\epsilon}\left(2C_A-\beta_{0l}\right)\left.T_{\mathrm{LO}}(x,\xi;q)\right|_{D=4},
\label{eq_gluon_z3_ir}
\end{equation}
where $\beta_{0l}=11C_A/3-4T_Fn_l/3$, $C_A=3$, $T_F=1/2$, and $n_l$ is the number of light-quark flavors.
As in the light-quark channel, GPD renormalization in the $\overline{\mathrm{MS}}$ scheme introduces a counterterm, which now involves the gluon evolution kernel:
\begin{equation}
T_{\mathrm{NLO,GPD}}^g=\frac{\alpha_s}{2\pi\epsilon}\int_{-1}^{1}\mathrm dy\,\left.T_{\mathrm{LO}}(y,\xi;q)\right|_{D=4}V_{gg}^{(0)}(y,x;\xi).
\label{eq_gluon_gpd_pole}
\end{equation}
Evaluating the convolution and combining it with Eqs.~(\ref{eq_gluon_coll_pole}) and (\ref{eq_gluon_z3_ir}), we find
\begin{equation}
T_{\mathrm{NLO,coll}}^g+T_{\mathrm{NLO},Z_3}^g+T_{\mathrm{NLO,GPD}}^g=0.
\label{eq_gluon_coll_cancellation}
\end{equation}

We next consider the soft contributions, which can be divided into three classes according to the subgraphs connected by the soft gluon.

\textit{Heavy-quark--antiquark exchange.}
This case is exemplified by the red line in Fig.~\ref{fig_feynNLO_g}(c).
We denote the sum of the soft poles from these exchanges by $T_{\mathrm{hh,soft}}$.
Combining it with the infrared contribution from on-shell heavy-quark field renormalization, we obtain
\begin{equation}
\left\langle T_{\mathrm{hh,soft}}+T_{\mathrm{NLO},Z_2}^g\right\rangle_S=\frac{2C_F\alpha_s}{3\pi\epsilon}\frac{\boldsymbol q^{\,2}}{E^2}T_{\mathrm{LO}}^{(0)}+\mathcal O(v^4),
\label{eq_gluon_hh_matching}
\end{equation}
where $T_{\mathrm{NLO},Z_2}^g=-C_F\alpha_s/(2\pi\epsilon)T_{\rm LO}$ and
$T_{\mathrm{LO}}^{(0)}$ denotes the Born amplitude at vanishing relative momentum.
This logarithmic infrared pole is reproduced by the ultraviolet-renormalized perturbative NRQCD matrix element~\cite{Bodwin:2008vp} and cancels in the matching.
The Coulomb singularity, which arises from the potential region, is not included in the expression above.
It is also reproduced by perturbative NRQCD and cancels in the matching~\cite{Bodwin:1994jh,Beneke:1997zp}.

\textit{Exchange between target-collinear gluons.}
This class is illustrated by the red line in Fig.~\ref{fig_feynNLO_g}(a).
With suitable loop-momentum assignments and $h$ denoting the soft-gluon momentum, the summed integrand admits the decomposition
\begin{equation}
I(h;q)=\frac{(n_+\cdot h)\mathcal A(h;q)+h_\perp^\mu h_\perp^\nu\mathcal B_{\mu\nu}(h;q)}{Q_0Q_aQ_b},
\label{eq_gluon_target_soft_suppression}
\end{equation}
where $\mathcal A$ and $\mathcal B_{\mu\nu}$ are analytic near $h=0$.
The three propagator denominators are
\begin{equation}
Q_0=h^2+i0,\quad
Q_a=h^2+2cE(n_+\cdot h)+i0,\quad
Q_b=h^2-2(1-c)E(n_+\cdot h)+i0.
\end{equation}
The soft energy poles of $Q_a$ and $Q_b$ lie on opposite sides of the real axis, allowing a Glauber pinch that prevents a direct application of the contour-deformation argument used in the light-quark channel.
Nevertheless, no infrared pole from this class survives as $\delta\to0$.

In the homogeneous soft limit, $h^\mu\sim\lambda E$, the denominators contribute a factor of $\lambda^{-4}$, while the two numerator terms are suppressed by at least $\lambda$ and $\lambda^2$, respectively.
Together with the $\lambda^4$ scaling of the integration measure in four dimensions, this gives a positive overall infrared degree.
Under the Glauber scaling, $h^0,h_z\sim\lambda^2E$ and $|\boldsymbol h_\perp|\sim\lambda E$, the denominators contribute a factor of $\lambda^{-6}$, while the numerator is suppressed by at least $\lambda^2$.
The integration measure scales as $\lambda^6$, so the overall infrared degree is again positive.

It remains to examine the target-collinear limits, $n_+\cdot h\sim\lambda^2\delta E$, $|\boldsymbol h_\perp|\sim\lambda\delta E$, and $n_-\cdot h\sim\delta E$, with $\delta$ held fixed as $\lambda\to0$.
Power counting allows logarithmic divergences in these limits.
Keeping the full quadratic denominators, we find that the collinear regions inside the soft sphere produce pole terms of order $\delta/\epsilon$.
These terms vanish as $\delta\to0$ and are neglected, consistently with our treatment in the collinear contribution.

\textit{Exchange between the target-collinear subgraph and the heavy-quark pair.}
The treatment of this class is similar to that in the light-quark channel.
In particular, a local deformation of the energy contour avoids the Glauber region and justifies the soft approximation while preserving the spatial cutoff.
Summing the relevant soft attachments, we obtain the singular term
\begin{align}
T_{\mathrm{NLO,soft}}^g={}&-\frac{i e g_s^4\mu_R^{2\epsilon}}{8\sqrt6\,(D-2)E^2(E+m_c)\xi^2}\frac{6-D+(D-2)r^2}{(1-r^2)^2}
\nonumber\\
&\times \int_{|\boldsymbol h|<\delta E}\frac{\mathrm d^Dh}{(2\pi)^D}\frac{(n_+\cdot h)(H\cdot J)-(H\cdot h)(n_+\cdot J)}{(h^2+i0)(n_+\cdot h-i0)^2}.
\label{eq_gluon_soft_current}
\end{align}
Here, $H$ and $J$ are defined in Eqs.~(\ref{eq_soft_hard_current}) and (\ref{eq_soft_dipole}), respectively.
Equation~(\ref{eq_gluon_soft_current}) includes both soft endpoints with the common spatial cutoff $\delta E$.

Evaluating the soft-momentum integral in Eq.~(\ref{eq_gluon_soft_current}), we obtain the residual infrared pole
\begin{equation}
T_{\mathrm{NLO,residual}}^g=\frac{e g_s^4}{2\sqrt6\,\pi^2\epsilon}\frac{x^2+\xi^2}{(x^2-\xi^2)^2}\frac{E\boldsymbol q_\perp^{\,2}+\frac{(q\cdot\varepsilon_\perp)^2}{E+m_c}\left[m_c(E+m_c)-\boldsymbol q_\perp^{\,2}\right]}{(E^2-q_z^2)^2}.
\label{eq_gluon_residual_pole_unaveraged}
\end{equation}
After averaging over the direction of $\boldsymbol q$ and expanding up to $\mathcal O(v^2)$ at fixed $E$, we obtain
\begin{equation}
\left\langle T_{\mathrm{NLO,residual}}^g\right\rangle_S=\frac{e g_s^4}{2\sqrt6\,\pi^2E}\frac{x^2+\xi^2}{(x^2-\xi^2)^2}\frac{1}{\epsilon}\left[\frac{\boldsymbol q^{\,2}}{E^2}+\mathcal O(v^4)\right].
\label{eq_gluon_residual_pole}
\end{equation}
We have verified this result through a complete one-loop calculation.

The current structure in Eq.~(\ref{eq_gluon_soft_current}) exhibits the same mechanism as in the light-quark channel.
At leading order in $v$, where the quark and antiquark have identical velocities, the heavy-quark dipole current $J$ vanishes, and so does the soft contribution; 
at order $v^2$, the terms linear in $q$ in $H$ and $J$ combine to produce the residual pole.
As discussed for the light-quark channel, no additional local vacuum-to-$J/\psi$ NRQCD matrix element is available to absorb this pole at this order.

\section{Summary and discussion}
In this work, we investigate the infrared structure of exclusive $J/\psi$ photoproduction up to relative order $\alpha_s v^2$ within the GPD and NRQCD factorization frameworks.
By analyzing the singular loop-momentum regions, we extract the soft and collinear singular terms and find nonvanishing infrared poles in both the gluon and light-quark channels after QCD renormalization, collinear factorization into the GPDs, and NRQCD matching.
We have verified the pole coefficients through a complete one-loop calculation.

The residual poles originate from incomplete cancellation of soft-gluon exchanges between the heavy-quark pair and the target-collinear partons.
The soft dipole current and the relevant hard current each contain a term linear in the relative momentum, whose product contributes at order $v^2$.
We examine the local NRQCD operator basis and find no additional independent vacuum-to-$J/\psi$ matrix element that can absorb these poles.
One possible extension for this process is to introduce nonlocal operators containing the target-collinear fields and NRQCD heavy-quark fields.
Whether operator mixing in such an extended framework can absorb the residual poles deserves further investigation.

Similar soft correlations may arise in other exclusive quarkonium production processes involving additional external QCD partons in the perturbative matching amplitude --- for an exclusive final state, one cannot rely on KLN cancellation through an inclusive sum over additional soft radiation.
An example is exclusive double-quarkonium production, where infrared divergences from incomplete cancellation of soft-gluon exchange between the constituents of the two quarkonia have been discussed in Ref.~\cite{Bodwin:2010fi}.
In that case, the corresponding contributions are suppressed by powers of the heavy-quark mass over the collision energy, whereas the residual poles found here exhibit no analogous suppression by powers of $m_{J/\psi}/W$.
For processes such as $\gamma\gamma\to J/\psi+\gamma$, the partonic matching amplitude contains no external QCD partons other than the heavy-quark pair.
At higher orders in $\alpha_s$, however, soft gluons can connect the heavy quarks to internal massless propagators.
Whether the corresponding loop-momentum regions generate analogous residual poles at relative order $\alpha_s^2v^2$ or beyond is another question worth investigating.

\vspace{1.0cm} 
{\bf Acknowledgments}
This work was supported by the National Natural Science Foundation of China (NSFC) under the Grants No. 12205061, 12622504. The work of LBC was also supported by the Guangdong Basic and Applied Basic Research Foundation under grant No. 2025B1515020009.
\vspace{1.0cm}

\end{document}